# Machine Learning for Classification of Protein Helix Capping Motifs

Sean Mullane, Ruoyan Chen, Sri Vaishnavi Vemulapalli,
Eli J. Draizen, Ke Wang, Cameron Mura, Philip E. Bourne

Data Science Institute; Department of Biomedical Engineering;
University of Virginia; Charlottesville, VA 22908 USA
{spm9r,rc3my,sv2fr,ed4bu,kw5na,cmura,peb6a}@virginia.edu

*Abstract*—The biological function of a protein stems from its 3-dimensional structure, which is thermodynamically determined by the energetics of interatomic forces between its amino acid building blocks (the order of amino acids, known as the *sequence*, defines a protein). Given the costs (time, money, human resources) of determining protein structures via experimental means such as X-ray crystallography, can we better describe and compare protein 3D structures in a robust and efficient manner, so as to gain meaningful biological insights? We begin by considering a relatively simple problem, limiting ourselves to just protein secondary structural elements. Historically, many computational methods have been devised to classify amino acid residues in a protein chain into one of several discrete "secondary structures", of which the most well-characterized are the geometrically regular α-helix and β-sheet; irregular structural patterns, such as 'turns' and 'loops', are less understood. Here, we present a study of Deep Learning techniques to classify the loop-like end cap structures which delimit α-helices. Previous work used highly empirical and heuristic methods to manually classify helix capping motifs. Instead, we use structural data directly—including (i) backbone torsion angles computed from 3D structures, (ii) macromolecular feature sets (e.g., physicochemical properties), and (iii) helix cap classification data (from CAPS-DB)—as the *ground truth* to train a bidirectional long short–term memory (BiLSTM) model to classify helix cap residues. We tried different network architectures and scanned hyperparameters in order to train and assess several models; we also trained a Support Vector Classifier (SVC) to use as a baseline. Ultimately, we achieved 85% class-balanced accuracy with a deep BiLSTM model.

*Index Terms*—bi-directional LSTM; Deep Learning; protein structure; secondary structure; alpha-helix; helix capping

## INTRODUCTION

Proteins, which are one of the basic types of biological macromolecules (along with the nucleic acids DNA and RNA), consist of long, polymeric chains of amino acid residues [1]. The vast array of known protein functions includes structuring and organizing the cell, binding and transporting molecules, transducing signals, and catalyzing biochemical reactions (enzymes). The detailed function of a protein is tied to its 3-dimensional structure and dynamics, and is ultimately governed by the statistical mechanics of the forces (atomic interactions) between its constituent amino acid residues; in general, the 3D structure is uniquely determined by the amino acid sequence.

Misfolding or alterations in protein structure and dynamics often compromise normal cellular functions, and such disruptions are often linked to disease states (hereditary and otherwise). For example, proteopathic diseases such as prion (e.g. Creutzfeldt-Jakob disease) and various amyloidoses are considered diseases of protein folding and aggregation. By understanding how sequences fold into unique structures, we can broaden our understanding of protein function and potentially develop novel medical treatments. At present, the 3D structures of only ~30% of human proteins have been empirically determined [2]. Thus, the problem of computationally *predicting* the 3D structure of a protein, given the amino acid sequence, remains as a grand challenge in biology.

Partitioning the backbone of a peptide chain (or really any biopolymer) into discrete, well-defined geometric segments is a first step of many common algorithms and workflows in structural bioinformatics, e.g. for the comparison and analysis of 3D protein structure. Many such approaches, while effective, often include at least one crucial step that is highly subjective in nature; for instance, the cutoff values for the $\phi/\psi$ values of a Ramachandran map may lead to the exact border of loop regions varying across algorithms, or even different implementations of the same structural assignment algorithm. A statistically well-principled, objective, and minimally heuristic segmentation approach would be a boon towards the development of improved automated methods for structural analysis, and would also provide a layer of abstraction (coarsening) from the peptide chain that could facilitate robust (quantitative) comparisons of structures across large databases (on genomic levels), with less computational expense. In addition, such an approach would bear upon quantitative definition of the 'fold' of a protein (and other biopolymers, such as RNA, where the problem of structural classification and comparison is even more difficult). As an initial sub-problem in this broader field, we consider the task of predicting helix capping motifs. In particular, we demonstrate that machine learning models can be trained to predict helix cap positions, offering a first step towards more robust and statistically-based classification systems for protein structural elements.

PRIOR WORK

Prior work on helix capping has used heuristic, highly manual approaches to study the residues in the cap regions that define the N- and C-termini of α-helices (these residues are termed 'Nt' and 'Ct', respectively). Historically, these cap regions have been grouped with "loop"-type secondary structures; however, in light of findings that their hydrophobic interactions are critical to the stability of helices and to super-secondary structural groups (e.g. α-α) [3], further subclassification may be warranted. Therefore, Aurora & Rose [3] described seven distinct capping motifs. More recent work used a density-based clustering approach—based on geometric quantities called the 'D' and 'delta' values of a cap, and the distribution of φ/ψ backbone torsion (i.e., Ramachandran) values—to identify 905 distinct clusters [4]. In another recent effort, 3D conformational clustering was used to map helix cap side-chain motifs to the loop structures that they support; this ability could be useful in de novo protein synthesis [5].

A common approach to protein secondary structure prediction is to use non-sequential models, typically feed-forward neural networks or SVMs [6]. However, these models (i) are not ideal for classifying data which cannot be naturally represented as a vector of fixed dimensionality, and (ii) cannot capture long-range dependencies in the data. Several authors have demonstrated the use of bidirectional LSTM networks for prediction of protein secondary structures and protein homology detection respectively ([7], [8]). Also recently, a feed-forward neural network has been trained to use NMR chemical shift information in order to predict structural motifs such as helix caps at the N-terminus, C-terminus and five types of β-turns [9].

DATA PRE-PROCESSING

Three data sources were chiefly used in this project. The first data source consists of all known protein structures from the full Protein Data Bank (PDB), in "macromolecular transmission format" (MMTF) format. The PDB has nearly 150,000 3D biomolecular structures, determined via experimental methods such as X-ray crystallography, nuclear magnetic resonance (NMR) spectroscopy, and cryo-electron microscopy. For all protein entries, we extracted secondary structural information (protein identifier, chain, residue type, and torsion angles.). The PDB does not contain information about helix caps, so we used CAPS-DB [4]—a relational database containing 67,530 helix caps (across 7390 proteins)—as the source of ground truth for the data we extracted from the PDB. In order to label the data for supervised learning, we merged the two datasets so that only proteins with known helix capping information were preserved. Based on capping information shown in Figure 1, residues in between the "startcap" and "endcap" positions were labeled as [0,1], to indicate helix caps. All other residues were given labels of [1,0]. We used this "one-hot encoding" of the binary variable so that we could use a Softmax output layer activation. The data pre-processing was done using the MMTF PySpark environment, which is a Python package that contains APIs for distributed analysis and scalable mining of 3D biomacromolecular structures, such as available from the PDB archive [10].

| pdbid | chain | type | start | end | startcap | endcap |
|-------|-------|------|-------|-----|----------|--------|
| 4kyc  | A     | Nt   | 66    | 72  | 66       | 66     |
| 2vch  | A     | Nt   | 123   | 129 | 123      | 123    |
| 4g7s  | A     | Nt   | 205   | 212 | 205      | 207    |
| 3ezi  | A     | Nt   | 107   | 116 | 107      | 109    |
| 4quk  | A     | Nt   | 99    | 117 | 99       | 101    |

FIGURE I
AN EXCERPTED SAMPLE FROM CAPS-DB

A major challenge in working with multiple types of data, generally from multiple sources (and with multiple degrees of standards compliance, if indeed there are standards at all), is ensuring that the data sets are merged correctly; indeed, "data wrangling" is such a major activity, from the perspective of software engineering in computational biology (and beyond), that entire books are dedicated to the topic [11]. Although the protein and protein chain IDs both used the standard PDB identification labels, the CAPS-DB data source utilizes a different resource, known as UniProtKB [12][13] for residue numbering. Therefore, we developed an intermediate "mapping layer" to bring the residue numbering schemes into proper correspondence. We used files downloaded from SIFTS [12][13] to determine the PDB ↔ UniProtKB mapping. Although this provided the correct residue numbering in the majority of cases, some mapping information was found to be missing in 1,962 chains out of 6,714, accounting for ≈1.7% of all residues and including 490 entire chains. In cases where residues were missing from a mapping file, we defaulted to labeling those residues as "not cap", the rationale being that this is (by far) the majority class and so, given no information, is more likely to be the correct label.

FEATURE ENGINEERING

From the PDB MMTF files we extracted the residue identity and atomic positions in 3D space, from which we calculated the φ and ψ backbone torsion angles. In CAPS-DB, cappings were extracted from high-quality protein structures and structurally clustered based on geometry and backbone conformation (i.e., φ/ψ values). Because the conformation of a given cap is encoded by a string of characters, each of which describes a precise region of the (φ/ψ)-space, torsion angles are among the most important features to consider in predicting helix cap positions [4]. Because neural networks typically train better when input features are distributed in roughly the range of a standard normal distribution, we normalized the torsion angles. This was done losslessly by using both the sine and cosine values of each torsion angle; in this way an angular value lying within [-180°, 180°] can be precisely represented, in a normalized form, by a pair of values, each lying between [-1, 1]. The four angular features and twenty one-hot encoded residue features comprised our basic, 24-feature data set.

We created a second dataset with additional features using a third data source, FEATURE [14]. This software computes various atomic features, as physicochemical descriptors, for each atom within each residue in an input protein structure. FEATURE calculates a quantized measure of these properties for the six concentric shells around each atomic site. The combination of six shells for each of the nine properties results in 54 additional features. In order to coarsen this data representation (from the atomic to the residue level), we took the maximum value of each feature, considered across all atoms in a given residue. Our basic feature set (i.e., the 24 descriptors mentioned above), combined with these additional properties, yields a total of 78 features.

TABLE I
RESIDUE CLASS FEATURES (PER SHELL)

| Class Number | Residue Class |
|---|---|
| 1 | IS_HYDROPHOBIC |
| 1 | IS_CHARGED |
| 1 | IS_POLAR |
| 1 | IS_UNKNOWN |
| 2 | IS_NONPOLAR |
| 2 | IS_POLAR |
| 2 | IS_BASIC |
| 2 | IS_ACIDIC |
| 2 | IS_UNKNOWN |

In terms of data preparation for the SVC model, a new dataset was generated by adding neighboring rows as features to each residue so as to account for the sequential nature of the data. A 'window size' parameter was used to determine the number of rows that would be added as features. For residues at the protein termini, rows before the residue were considered. In case of shorter sequences, for residues in the center, rows before and after it were considered. Sequences with length less than the window size were removed from the original data before these operations; this is done to avoid padding the data for sequences that are shorter than the window size. As a representative example, a window size of 10 resulted in this filtering operation removing 18 protein chains. Because of the added features (790 per residue), we used principal component analysis (PCA) to reduce the dimensionality of the feature space to 40 (accounting for 67.7% of variance in the data). Because the computational complexity of an SVC model with a nonlinear kernel (e.g., the radial basis function) is supra-quadratic ($O(n_{features}*n_{samples}^2)$), we sampled 10% of the data rows (98,078 samples spanning 3,968 protein chains) and divided into train and validation sets in a ratio of 70:30.

For the deep neural network models, we kept the protein chains intact as sequential data for the LSTMs. We used a random 80/20 training/validation split, separated at the protein chain level. We stored these as separate Pickle [15] files for predictors, labels and (for the encoder/decoder model) lagged labels.

MACHINE LEARNING MODELS

As a baseline approach for predicting helix caps, we trained a supper-vector clustering (SVC) model using a Radial Basis Function (RBF) kernel, with low γ and C values to get a simpler decision function [16].

For streams of data which are inherently sequential in nature (e.g., time-series), Recurrent Neural Networks (RNN) are frequently employed. These networks feed the hidden state of one timestep as an input to the next timestep. Recurrent networks are some of the most widespread deep learning techniques and are particularly effective for data which has sequential information with cyclic connections. This family of models includes the Long Short-Term Memory (LSTM) and Gated Recurrent Units (GRU), among others. LSTMs [17] are particularly useful when there are long-range dependencies within the sequential data; they can automatically detect both the long-term and short-term dependency relationships, and determine how to process a current subsequence according to the information extracted from the prior subsequences [8]. Because proteins are generally quite compact, globular entities—with chains that can tightly fold back upon themselves—they do indeed exhibit a network of long-range dependencies: the string of amino acids defines the residue sequence, and residues far apart in the linear sequence are often proximal in 3D space. Compared to other methods, LSTMs can better identify patterns of protein homology in purely sequence-based data [8]. In our study, the primary model was a bidirectional–LSTM model. The bidirectional layer, added to a normal LSTM, allows the model to deal with both forward and backward dependencies and enhances predictive performance [18].

MODEL OPTIMIZATION AND EVALUATION

To evaluate our models, we used three measures of performance: accuracy, balanced accuracy (Qα) (1) , and F1 score (2).

$$Q_\alpha = \frac{1}{2} * (precision + recall) \quad (1)$$

$$F1 = 2 * \frac{(precision*recall)}{(precision+recall)} \quad (2)$$

Precision and recall are defined as:

$$precision = \frac{TN}{TN+FP} \quad (3)$$

$$recall = \frac{TP}{TP+FN} \quad (4)$$

In a binary classification framework, assuming equal weight per residue and independence among residues, the performance of a classifier can be fully described by the four values True Negative, False Negative, True Positive and False Positive [19]. No single measure can perfectly distill this information, but in light of the class imbalance in our problem (namely, that ~85% of all residues in our dataset are not caps), we primarily consider the Qα and F1 metrics; these two quantities account for the class imbalance, whereas the raw accuracy (a single-class measure) can be misleadingly high, even in the case of completely uninformative (i.e., random) predictions, simply by virtue of the class imbalance.

Although evaluating protein structure prediction methods can be confounded somewhat when homologous sequences occur in the data set, for the sake of simplicity we did not try to account for effects of protein homology in this work. To mitigate overfitting on particular protein chains or chain families in the training data set, our neural network architecture employed regularization [20]. Also, note that we took each residue in a helical cap as being equally informative in weighting for our training loss function and evaluation metrics. In future work, it may be helpful to consider some parts of a cap, e.g. the central Ct and Nt residues, as being more informative than, e.g., the last residue of the cap; potentially, this could be achieved by encoding as a feature various quantities such as the information content of each position (number of bits, e.g. in a profile hidden Markov model).

We examined several different BiLSTM model architectures in order to try to find the optimum classifier performance with our datasets. The variations included (i) single-layer as well as deep LSTMs, (ii) the number of nodes in the BiLSTM hidden layers, (iii) the weighting of the loss function, (iv) the addition of dropout layers with varying drop ratios, and (v) single as well as multiple dense layers after the BiLSTM layers, with and without activation functions. We implemented these models using Keras 2.2.4 with a TensorFlow 1.12.0 backend.

Each model was trained using cross-entropy loss functions and using the Adam optimizer, which is a form of stochastic gradient descent [21]. Training periods varied from 30 to 75 epochs. We trained the models with a batch size of unity. The rationale for this was that, because of the varying sequence lengths in the data, batched training would require padding to equalize these lengths. However, technical limitations in the CuDNNLSTM layer in Keras [22], which provides a GPU-enabled library for deep neural networks, would cause the padded values to be considered in training and evaluation (this layer does not support masking), and that would be problematic for both training and evaluation.

In addition to varying the basic model architecture itself, we also trained models with both the smaller set of 24 features and the expanded set of 78 features, so as to determine the relative utility of the additional features. The results of our calculations are summarized in the remainder of this work.

TABLE II
MODEL ARCHITECTURE

| Run | Feats. | LSTM Layers | LSTM Nodes | Dense Layers | Dense Activ. | Weight Ratio | Dropout % |
|---|---|---|---|---|---|---|---|
| 1 | 78 | 1 | 100 | 1 | NA | 1:1 | 0 |
| 2 | 78 | 1 | 100 | 1 | NA | 8:1 | 0 |
| 3 | 78 | 1 | 100 | 1 | NA | 8:1 | 25 |
| 4 | 78 | 1 | 100 | 1 | NA | 8:1 | 50 |
| 5 | 24 | 1 | 100 | 1 | NA | 8:1 | 0 |
| 6 | 24 | 1 | 20 | 1 | NA | 8:1 | 0 |
| 7 | 24 | 1 | 40 | 1 | NA | 8:1 | 0 |
| 8 | 24 | 1 | 40 | 1 | NA | 8:1 | 50 |
| 9 | 24 | 1 | 40 | 1 | NA | 7:1 | 50 |
| 10 | 24 | 1 | 40 | 1 | NA | 6:1 | 50 |
| 11 | 24 | 1 | 40 | 1 | NA | 5:1 | 50 |
| 12 | 24 | 1 | 40 | 2 | ReLU | 8:1 | 50 |
| 13 | 24 | 2 | 40,20 | 2 | ReLU | 8:1 | 50 |
| 14 | 24 | 3 | 40,30,20 | 2 | ReLU | 8:1 | 50 |
| 15 | 78 | 2 | 100,50 | 1 | NA | 8:1 | 50 |
| 16 | 78 | 2 | 100,50 | 2 | None | 8:1 | 50 |
| 17 | 78 | 2 | 40,10 | 1 | NA | 8:1 | 50 |

RESULTS

The SVC model did not perform well, as might be expected. The class-balanced accuracy was only 50.25%, and the recall values indicate that the predictions were nearly single-class and thus nearly devoid of information. The model's deficiency may stem from its inability to capture long-ranged patterns/dependencies in the data—e.g., between amino acid residues that are quite distant in sequence, but potentially correlated (near one another) in 3D space.

TABLE III
SVC MODEL PERFORMANCE

| Class | Precision | Recall | F1 score |
|---|---|---|---|
| Not Cap | 0.87 | 0.99 | 0.93 |
| Cap | 0.72 | 0.01 | 0.01 |

Of the BiLSTM models, the best-performing overall was a deep model built with three BiLSTM layers (with 40, 30, and 20 hidden nodes), followed by two dense layers with 50% dropout between each layer. The first of these dense layers used nodes with rectified linear unit (ReLU) activation functions, and it had 10 hidden nodes.

TABLE IV
BiLSTM MODEL PERFORMANCE, BY BEST VALIDATION SET F1 SCORE

| Run | Accuracy | F1 | $Q_\alpha$ | Best Epoch |
|---|---|---|---|---|
| 1 | **0.8764** | 0.4791 | 0.6896 | 11 |
| 2 | 0.8155 | 0.5228 | 0.7907 | 12 |
| 3 | 0.8132 | 0.5212 | 0.8000 | 6 |
| 4 | 0.8034 | 0.5182 | 0.8078 | 8 |
| 5 | 0.8155 | 0.5228 | 0.7907 | 12 |
| 6 | 0.8188 | 0.5392 | 0.8264 | 36 |
| 7 | 0.8262 | 0.5489 | 0.8297 | 11 |
| 8 | 0.8213 | 0.5454 | 0.8318 | 15 |
| 9 | 0.8219 | 0.5429 | 0.8273 | 26 |
| 10 | 0.8369 | 0.5535 | 0.8202 | 13 |
| 11 | 0.8504 | **0.5612** | 0.8096 | 15 |
| 12 | 0.8030 | 0.5321 | 0.8376 | 12 |
| 13 | 0.8145 | 0.5480 | **0.8457** | 48 |
| 14 | 0.8234 | 0.5551 | 0.8430 | 31 |
| 15 | 0.8087 | 0.5332 | 0.8223 | 11 |
| 16 | 0.8096 | 0.5254 | 0.8103 | 31 |
| 17 | 0.8070 | 0.5236 | 0.8109 | 41 |

In general, we found improved performance across our metrics with each additional layer; we did not test models deeper than those described above. The best performance was achieved with those models that used an initial hidden layer with a node count moderately larger than the feature count; significantly more hidden nodes than features resulted in overfitting, while fewer resulted in lower overall accuracy statistics.

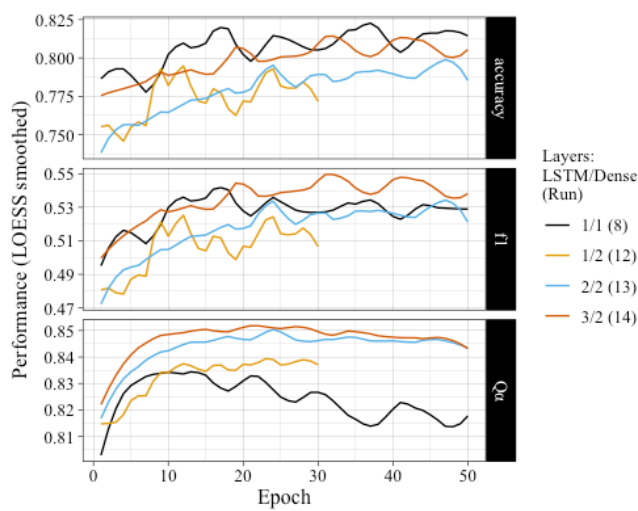

FIGURE II
EFFECT OF NEURAL NETWORK LAYER DEPTH

The loss function weighting scheme had a large effect, as might be anticipated. The models trained with the un-weighted binary cross-entropy loss were found to heavily optimize for overall accuracy, at the expense of recall; this result is reflected in the lower $Q\alpha$ and F1 scores. When using a cross-entropy loss with a class weighting inversely proportional to the prevalence of each of the two classes (cap, not cap), the $Q\alpha$ and F1 scores were found to be maximal. However, we observed a trade-off in terms of the $Q\alpha$ and F1 scores, as we varied the class weight ratio from the 8:1 to 5:1; the latter had a higher F1 score while the former had higher $Q\alpha$ as it more severely over-predicted the cap class compared to the training set prevalence.

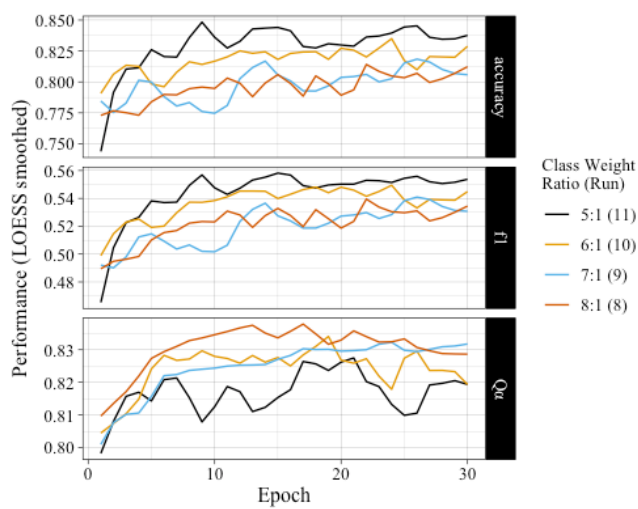

FIGURE III
EFFECT OF NEURAL NETWORK LOSS FUNCTION WEIGHTING

Inclusion of additional features, as derived from the FEATURE code, were not found to improve the predictive performance of our models. This is a surprising result, as hydrophobicity and polarity are believed to be important physicochemical factors in helix capping interactions. Our 'aggregation' method (taking the maximal atom-based value per residue and mapping that to the entire residue) was potentially suboptimal; future work can consider other schemes for combination/aggregation of these numerical values.

## CONCLUSION

Our results reveal that the residue positions of helix caps—or at least some sort of non-random (detectable) structural 'signature', which we take as being helix caps—can be learned purely from amino acid sequences along with corresponding conformational data (backbone $\varphi/\psi$ angles). This finding demonstrates that machine learning can be used to identify patterns of helix capping, and suggests the possibility of discovering a statistically robust, objective cap classification scheme (i.e., minimal heuristics). We also find that bidirectional LSTMs are particularly well-suited to classifying protein sequences, likely because of the importance of both long-term and short-term atomic interactions in dictating protein folding and structural stability. In future work, we could focus on developing an encoder/decoder sequence-to-sequence BiLSTM model, as such approaches offer superior performance (versus standard BiLSTM models) in some scientific domains.

In terms of broader biomedical relevance, we note that helices are viewed by many protein biophysicists as being less prone to aggregation, versus β-rich elements; see, for instance, various reviews in the extensive amyloid/protein aggregation literature (e.g., [23]). Indeed, it has been argued that sequestering sequences with β-propensities into helices may offer a way to avoid aggregation [24]. As a potentially interesting future direction, we can consider questions such as whether engineering helix-capping signals into particular disease-associated (β-forming) peptide regions (e.g., via the CRISPR technology) may be a way to reduce their amyloidogenicity? In this sense, note that studies of protein helix capping are potentially of both fundamental significance as well as more applied relevance.

## ACKNOWLEDGMENT

We thank P. Alonzi (UVA) for AWS support. UVA's Open Data Lab and Data Science Institute provided computational resources that contributed to our results. Portions of this work were also supported by UVA and NSF CAREER award MCB-1350957.

## REFERENCES

[1] Kuriyan, J., Konforti, B. and Wemmer, D. The Molecules of Life: Physical and Chemical Principles. illustrated. Garland Science.
[2] Trevizani, R., Custódio, F.L., Dos Santos, K.B. and Dardenne, L.E. 2017. Critical features of fragment libraries for protein structure prediction. Plos One 12(1), p. e0170131.

[3] Aurora, R. and Rose, G.D. 1998. "Helix capping." Protein Science 7(1), pp. 21–38.

[4] Segura, J., Oliva, B. and Fernandez-Fuentes, N. 2012. "CAPS-DB: a structural classification of helix-capping motifs." Nucleic Acids Research 40(Database issue), pp. D479-85.

[5] Newell, N.E. 2015. "Mapping side chain interactions at protein helix termini." BMC Bioinformatics 16, p. 231.

[6] Hua, S. and Sun, Z. 2001. "Support vector machine approach for protein subcellular localization prediction." Bioinformatics 17(8), pp. 721–728.

[7] Sønderby, S.K. and Winther, O. 2014. "Protein Secondary Structure Prediction with Long Short Term Memory Networks." arXiv.

[8] Li, S., Chen, J. and Liu, B. 2017. "Protein remote homology detection based on bidirectional long short-term memory." BMC Bioinformatics 18(1), p. 443.

[9] Shen, Y. and Bax, A. 2012. "Identification of helix capping and b-turn motifs from NMR chemical shifts." Journal of Biomolecular NMR 52(3), pp. 211–232.

[10] "Structural Bioinformatics Training Workshop & Hackathon 2018" - github.com/sbl-sdsc/mmtf-workshop-2018

[11] "Principles of Data Wrangling O'Reilly Media" - shop.oreilly.com/product/0636920045113.do

[12] Velankar, S., Dana, J.M., Jacobsen, J., van Ginkel, G., Gane, P.J., Luo, J., Oldfield, T.J., O'Donovan, C., Martin, M.-J. and Kleywegt, G.J. 2013. "SIFTS: Structure Integration with Function, Taxonomy and Sequences resource." Nucleic Acids Research 41(Database issue), pp. D483-9.

[13] Dana, J.M., Gutmanas, A., Tyagi, N., Qi, G., O'Donovan, C., Martin, M. and Velankar, S. 2019. "SIFTS: updated Structure Integration with Function, Taxonomy and Sequences resource allows 40-fold increase in coverage of structure-based annotations for proteins." Nucleic Acids Research 47(D1), pp. D482–D489.

[14] Bagley, S.C. and Altman, R.B. 1995. "Characterizing the microenvironment surrounding protein sites." Protein Science 4(4), pp. 622–635.

[15] "Python object serialization" - docs.python.org/3/library/pickle.html

[16] "sklearn.decomposition.PCA" scikit-learn.org/stable/modules/generated/sklearn.decomposition.PCA.html

[17] Goodfellow, Ian, Bengio, Yoshua, and Courville, Aaron. 2016. "Deep Learning", MIT Press

[18] Cui, Zhiyong, Ruimin Ke, and Yinhai Wang. "Deep Stacked Bidirectional and Unidirectional LSTM Recurrent Neural Network for Network-wide Traffic Speed Prediction." 6th International Workshop on Urban Computing (UrbComp 2017). 2016.

[19] Baldi, P., Brunak, S., Chauvin, Y., Andersen, C.A. and Nielsen, H. 2000. "Assessing the accuracy of prediction algorithms for classification: an overview." Bioinformatics 16(5), pp. 412–424.

[20] Srivastava, N., Hinton, G., Krizhevsky, A., Sutskever, I. and Salakhutdinov, R. 2014. "Dropout: a simple way to prevent neural networks from overfitting." The Journal of Machine Learning Research 15(1), pp. 1929–1958.

[21] Kingma, D.P. and Ba, J. 2014. "Adam: A Method for Stochastic Optimization."

[22] "Keras" 2015. keras.io. Accessed: March 31, 2019.

[23] Eisenberg, D.S. and Sawaya, M.R. 2017. "Structural studies of amyloid proteins at the molecular level." Annual Review of Biochemistry 86, pp. 69–95.

[24] Tzotzos, S. and Doig, A.J. 2010. "Amyloidogenic sequences in native protein structures." Protein Science 19(2), pp. 327–348.

**AUTHOR INFORMATION**

**Sean Mullane,** Data Scientist, UVA Health System, M.S. Student, Data Science Institute, University of Virginia

**Ruoyan Chen,** M.S. Student, Data Science Institute, University of Virginia

**Sri Vaishnavi Vemulapalli,** M.S. Student, Data Science Institute, University of Virginia

**Eli J. Draizen,** PhD Student, Bourne & Mura Computational Biosciences Lab, Department of Biomedical Engineering, University of Virginia

**Ke Wang,** Research Scientist, Department of Computer Science, University of Virginia

**Cameron Mura,** Senior Scientist, Bourne & Mura Computational Biosciences Lab, Department of Biomedical Engineering, University of Virginia

**Philip E. Bourne,** Stephenson Chair of Data Science, Director of the Data Science Institute, and Professor of Biomedical Engineering, University of Virginia